\documentclass[proceedings]{stacs}
\stacsheading{2009}{13--30}{Freiburg}
\firstpageno{13}

\usepackage{epsfig}
\usepackage{amssymb}
\usepackage{amsmath}
\usepackage{amsfonts}
\usepackage{subfigure}

\begin{document}
\title{A Comparison of Techniques for Sampling Web Pages}
\author[lab1]{E. Baykan}{Eda Baykan}
\author[lab1,lab2]{M. Henzinger}{Monika Henzinger}
\author[lab3]{S.F. Keller}{Stefan F. Keller}
\author[lab3]{S. de Castelberg}{\\Sebastian de Castelberg}
\author[lab3]{M. Kinzler}{Markus Kinzler}
\address[lab1]{Ecole Polytechnique F\'ed\'erale de Lausanne (EPFL)  
  \newline IC LTAA Station 14 CH-1015 Lausanne Switzerland
}  
\email{eda.baykan@epfl.ch}  
\address[lab2]{Google Switzerland
}
\address[lab3]{University of Applied Science Rapperswil, Switzerland
}	
\thanks{Preliminary results of this paper were presented at IIWeb 2006 Workshop}

\begin{abstract}
	\noindent As the World Wide Web is growing rapidly, it is getting increasingly challenging to gather representative
information about it. Instead of crawling the web exhaustively one has to resort to other techniques like
sampling to determine the properties of the web. A uniform random sample of the web would be useful to
determine the percentage of web pages in a specific language, on a
topic or in a top level domain.
Unfortunately, no approach has been shown to sample the web pages in an  unbiased way.
Three promising web sampling algorithms are based on random walks.
They each have been evaluated individually, but making a comparison on different data sets is not possible.
We directly compare these algorithms in this paper.
We performed three random walks on the web under the same conditions and analyzed their outcomes in detail.
We discuss the strengths and the weaknesses of each algorithm and propose improvements
based on experimental results.
\end{abstract}

\maketitle

\section*{Introduction}
The World Wide Web is a rich source of information about the world but
very little information is known about the web itself. We do not
know what percentage of web pages are in a specific language or on a
topic or in a top level domain. There are estimates on
what percentage of web pages change per day~\cite{cho-garcia-molina-bk00,fetterly-et-al-toplas03} but
they depend on how deeply the sites
were crawled.  Trying to determine these statistics
based on exhaustive enumeration of the web is not feasible because of
its size and its rapidly changing nature.  However, a uniform random
sample of the web\footnote{We refer to a uniform random sample of the web as the uniform random sample of the web pages {\em not} of the graph structure of the web.} would provide answers to many of the above
questions and repeated sampling would also allow to monitor changes
in the web's composition. 

In the literature there are four major approaches for sampling web pages:
Lawrence and Giles~\cite{lawrance-giles-bk99}  tested random IP addresses to
determine characteristics of hosts. However, it leaves the question
open how to deal with multiple hosts sharing the same IP address or
hosts being spread over multiple IP addresses. Additionally, it is not
clear how to sample the web pages accessible at a given IP address. Thus, 
this approach samples IP addresses, but not web pages.

Bar-Yossef et al.~\cite{bar-yossef-et-al-toplas-00} and Henzinger et al.~\cite{henzinger-et-al-toplas99,henzinger-et-al-toplas00} 
independently proposed to use random walks on
the web to sample web pages. They present algorithms that in
theory should lead to uniform random samples but cannot be
implemented in their {\em pure} form.  Instead, the implementations need
to make some simplifications which lead to various biases in the
resulting samples.  Both evaluated their walks on different artificially generated graphs
and on the web (at different times).
Based on this work, Rusmevichientong et al.~\cite{nec} proposed two
different random walks, which in theory should lead to uniform random
samples. One of their approaches can be implemented {\em without modifications}. However, they
evaluated their approaches only on small artificially generated graphs
consisting of $100,000$ nodes. On these graphs they showed that their
approaches and the approach in~\cite{bar-yossef-et-al-toplas-00} lead
to samples that reflect the indegree and outdegree distributions of
the underlying graph correctly, while the approach by 
 Henzinger et al.~\cite{henzinger-et-al-toplas99,henzinger-et-al-toplas00} does not. 
Henzinger et al.~\cite{henzinger-et-al-toplas00} had found a bias in their approach for the indegree distribution but not
for the outdegree distribution. More recently, Bar-Yossef et al.~\cite{yossef-et-al-focusedsampling} showed how
to generate a random sample of web pages relevant to a given user specified topic and 
Chakrabarti et al. ~\cite{chakrabarti-et-al-02} developed techniques to estimate the background topic distribution
on the web.
Both ~\cite{yossef-et-al-focusedsampling} and ~\cite{chakrabarti-et-al-02} use a variant of the sampling
algorithm proposed in ~\cite{bar-yossef-et-al-toplas-00}.

In the rest of the paper we will denote 
the algorithm proposed in~\cite{nec} as {\em Algorithm~A}, the algorithm proposed in~\cite{bar-yossef-et-al-toplas-00} as {\em Algorithm~B} and
the algorithm in~\cite{henzinger-et-al-toplas00} as {\em Algorithm~C}. 
Each algorithm consists of a {\em walk phase} that performs a random walk
and of a {\em subsampling phase} that subsamples the web pages visited by the random walk.
We performed the walk phase of each of these algorithms {\em on the web} with the {\em same} computation power
and with the {\em same} amount of time. Then we experimented with different subsampling phases, including the ones
proposed by the above papers. This resulted in four types of samples generated by Algorithm~A, called {\em A~Samples}, 
four types of samples generated by Algorithm~B, called {\em B~Samples},
and three types of samples generated by Algorithm~C, called {\em C~Samples}.

Our experiments provide the following new insights about the above mentioned algorithms:  
(1) A~Samples and B~Samples exhibit
a strong bias to internally highly connected hosts with few outedges to other
hosts. The reason is that Algorithm~A and Algorithm~B frequently had problems leaving such hosts.
After a certain (large) number of consecutive visits of web pages on the same host, we say that the walk
is {\em unable to leave} the host or, more informally, got {\em stuck} at a host.
Both Algorithm~A and Algorithm~B have a problem with getting stuck.
Algorithm~C is designed to have a very low probability of getting stuck, due to {\em random resets.}
Indeed in our experiments it was never unable to leave a host.
(2) C~Samples exhibit a bias towards high outdegree web pages.
This was shown before for artificially generated graphs~\cite{nec} but not for the web.
Furthermore we show that C Samples show a bias towards high PageRank web pages.
(3) We experimented with different subsampling phases
for each algorithm. The subsampling techniques had
an impact on A~Samples and B~Samples while they had only a very small impact 
on C~Samples. 

This paper is organized as follows: Section~\ref{tab:Description}
describes the evaluated algorithms and their corresponding
subsampling phases.
Section~\ref{tab:details} presents some challenges met during the
implementation and how we dealt with them. 
Section~\ref{tab:experiments} presents the experiments
and their results in detail. In Section ~\ref{tab:summary} we
give a comparison of results for sampling algorithms. 
We conclude with proposals for further work in Section~\ref{tab:futureWork}.

\section{Description of the algorithms}\label{tab:Description}
We define the {\em web graph} as a graph where every web page is a node 
and every hyperlink is a directed edge between the nodes. A {\em memoryless random
walk} on the web graph is a Markovian chain that visits a sequence of
nodes where the {\em transitions} between nodes depend only on the last node of the walk and not
on earlier nodes. In a Markov chain on the web graph {\em states} correspond to web pages, 
i.e. the nodes on the web graph, and each visit to a node results in one {\em step} of the random walk. 
We call a step a {\em selfloop step} when the walk
visits the same node in two consecutive steps of the walk by traversing
a selfloop. We define the {\em visit count} of a node to be the number of visits to 
the node including selfloop steps. 
{\em Edges}, {\em degree}, {\em PageRank}, 
{\em inlinks}, {\em outlinks} and {\em selfloop} of a state are the edges, degree, PageRank
value, inlinks, outlinks and selfloop of the corresponding node.

Each algorithm consists of two phases:
(1) A {\em walk phase}, where a memoryless random walk is performed on the web graph.
We denote the walk phase of Algorithm~A, Algorithm~B and Algorithm~C as {\em Walk~A}, {\em Walk B} and {\em Walk~C} respectively.
(2) The second phase is a {\em subsampling phase}, where either states or steps of the walk phase are subsampled randomly.

According to a fundamental theorem of Markovian chains, a random walk
on an aperiodic and irreducible graph will converge to a unique stationary distribution.
Once the walk reaches its unique stationary distribution, the probability of being in a node
will not change although the walk takes more steps.
Algorithm~A and Algorithm~B are designed to perform a random walk on an undirected, aperiodic and irreducible
graph. On such a graph a random walk converges to a unique stationary distribution
where the probability of being in a node is proportional to its degree.
Walk~A leads to a biased stationary distribution because the nodes do not have the same degree.
If we subsample states after the point where the walk reaches the stationary distribution,
high degree nodes will be more likely to be sampled. 
To remove this bias we subsample states or steps of Walk~A with 
values inversely proportional to the corresponding node's degree.
Walk B is performed on a regular graph, i.e. on a graph where each node has the same degree.
Furthermore this regular graph has the above mentioned properties required
for converging to a unique stationary distribution. Thus, in
the stationary distribution values of Walk B every node is equally likely to be visited.
Algorithm~C is designed to perform a random walk on a directed, aperiodic and irreducible
graph. This walk leads to a unique stationary distribution where the probability of being in a node is
equal to its PageRank value. In other words
Walk~C has a biased stationary distribution, as does Walk~A.
To get a uniform random sample of Walk~C,
we subsample its {\em states} with values inversely proportional
to the PageRank values of the corresponding nodes.
We next describe the algorithms in more detail. 

\subsection{Algorithm~A}\label{tab:DescriptionAlgoritmA}
{\em Walk phase:}
Consider the following random walk on an {\em undirected} graph. From the current node
choose an adjacent edge uniformly at random and select the other endpoint of the
edge as next node to visit. It can be proven that if run long enough 
on an {\em undirected, irreducible and aperiodic} graph this random walk converges to a 
unique stationary distribution where the probability of visiting a node is proportional 
to its degree.

Algorithm~A executes Walk~A on the web graph which it modifies as follows: 
(1) It gives a selfloop to each node that does not yet have a selfloop to make the
web graph aperiodic. (2) It ignores the direction of the (directed) hyperlinks. The latter
leads to complications in the implementation since the inlinks of a node
in the web graph can not be determined directly from the corresponding web page. Additionally the web graph
changes constantly as web pages are edited. We deal with the former problem by querying a web search engine
and retrieving up to 10 inlinks per node, chosen randomly from all returned inlinks. We do not
retrieve more inlinks since it was shown experimentally in~\cite{bar-yossef-et-al-toplas-00} that Algorithm~A
returns better results when the number of retrieved inlinks is limited. These inlinks and outlinks together
with the inlinks (if any) from previously visited nodes form the set of adjacent edges of a node.
To deal with the changes in the web we store the set of adjacent edges of a node at the first visit of
the node in a database. At every later visit of the node the set of its adjacent edges is taken
from the database. This guarantees that the degree of a node does not change
during the execution of the walk.

{\em Subsampling phase:}
After Walk~A reaches its unique stationary distribution each node can be the next step of the walk
with probability proportional to its degree. To remove this bias, states or steps are subsampled randomly
with probability inversely proportional to their degree after
the step where the walk reached the stationary distribution.
We wanted to implement the algorithms described in~\cite{bar-yossef-et-al-toplas-00} and ~\cite{nec}
as closely as possible, however it was not clearly described whether they subsampled states or steps. Thus
we created two types of samples, one subsampling states and one subsampling steps.
The number of steps until the walk has reached a stationary distribution is
called the {\em mixing time.} No bounds for the mixing time on the web graph are known.
However, intuitively the distribution
of states towards the end of the walk should lead to better results than the distribution of all the states 
in the whole walk. We tested this intuition by exploring the following different subsampling phases.
(1) We determined all the states visited in the last half of the steps of Walk~A and subsampled them 
randomly with probability inversely proportional to their degree.
This sample is called {\em A\_StatesOnLastHalf}.
(2) We determined all states visited in the last quarter of the steps of Walk~A and 
subsampled them randomly with probability inversely proportional to their degree.
This sample is called {\em A\_StatesOnLastQuarter}.
(3) From the last half of the steps of Walk~A, the steps are subsampled randomly with probability 
inversely proportional to the corresponding web page's degree. This sample is called {\em A\_StepsOnLastHalf}.
(4) From the last quarter of the steps of Walk~A, the steps are subsampled randomly with probability inversely proportional to the
corresponding web page's degree. This sample is called {\em A\_StepsOnLastQuarter}.

\subsection{Algorithm~B}
{\em Walk phase:}
Consider the same random walk as for Walk~A on an {\em undirected, regular and irreducible} graph.
If run long enough this random walk converges to a uniform distribution of the nodes.
The web graph is neither undirected nor regular.
The web graph is modified as described in Section~\ref{tab:DescriptionAlgoritmA}
to make it undirected.
To make it regular we add enough selfloops to each
node to increase their degree to $max.$ 
Following ~\cite{bar-yossef-et-al-toplas-00} we set $max=10,000,000.$
Thus, the only difference between Walk~A and Walk~B is the number of selfloops in the graph.

{\em Subsampling phase:}
We subsampled Walk B in the same four ways as Walk~A, but subsampling states uniformly at 
random creating B\_StatesOnLastHalf, B\_StatesOnLastQuarter and
subsampling steps uniformly at random creating the samples B\_StatesOnLastHalf, B\_StepsOnLastQuarter.

\subsection{Algorithm~C}
{\em Walk phase:} Algorithm~C tries to imitate the PageRank random walk~\cite{brin-page-bk98} as closely as possible.
When choosing the next node to visit, Walk~C first flips a biased coin.
With probability $d = 1/7$ it performs a {\em random jump} or {\em random reset}, described below.
With probability $1-d$, it chooses an outlink of the current node uniformly at
random and selects the head of the selected outlink as next step of the walk. We say that the algorithm
{\em traverses} the chosen outlink.
If a chosen node does not have any outlinks or if it cannot be fetched, a random jump is performed.
Ideally a random jump would jump to a randomly selected node of the web graph. However, the walk
does not know all the nodes on the web graph. Instead it can choose a node out of all
{\em visited} or all {\em seen} nodes. A node is {\em seen} if it either has already been visited
or if it is the head of an outlink of a visited node. However, even when restricting
the random jumps to all visited or all seen nodes, there is a potential problem.
As pointed out in~\cite{henzinger-et-al-toplas99} if almost all of the seen nodes are on the same host, a random jump
would with high probability jump to a node on this host. As a result it is possible that the walk
gets stuck on this host.
To remedy this problem ~\cite{henzinger-et-al-toplas99} proposed to perform
a random jump in the following biased way: First select a host from all the visited
hosts uniformly at random, then select a web page from all the visited web pages on that host
uniformly at random and finally visit the node corresponding to the selected web page.
In our implementation we ``got stuck'' in domains using this approach and thus
we added one additional layer, the domain\footnote{We denote by {\em domain} second level domains 
like {\em epfl.ch} or {\em berkeley.edu}.} layer. Additionally we switched from {\em visited}
to {\em seen} entities. A {\em seen host} is a host on which the walk has seen a web page
and a {\em seen domain} is a domain on which the walk has seen a web page.
Our Walk~C first selects a domain uniformly
at random from all the seen domains, secondly it selects a host  uniformly  at random
from all the seen hosts in that domain, then it selects a web page uniformly at random from 
all the seen web pages in that host and finally the walk visits the node corresponding to the selected web page.
Since the set of seen nodes is on the average a factor of roughly
10 larger than the set of visited nodes, this modification allowed us to more closely imitate the
PageRank random walk that chooses a random node out of {\em all} nodes on the web graph in random jump phase.
Due to our way of imitating PageRank random walk our Walk~C is not memoryless since it keeps track of all the visited states 
as well as their outlinks.

{\em Subsampling phase:}
Following~\cite{henzinger-et-al-toplas00} we use three different subsampling phases to subsample states of
Walk~C. One is simply a uniform random sample of all the nodes, called {\em C\_Random}. 
However this will be biased towards high PageRank nodes as they are more likely to be visited. The other
two sampling techniques try to correct for this bias. The idea is to sample
inversely proportional to PageRank values. Since the PageRank values for the whole web is not
known, a {\em PageRank substitute} is used during the sampling. It
is computed in one of the two possible ways: (1) The PageRank of the subgraph of the visited states is 
computed and the visited states are subsampled inversely proportional to their PageRank values. 
This sample is called {\em C\_PR}.
(2) The ratio of the number of visits of a node to the total number of steps
of the walk is called as {\em visit ratio} of the node.
The PageRank random walk converges to a unique stationary distribution where the probability
that a node is visited is proportional to its PageRank value. In the limit, i.e. when the length of
the walk goes to infinity, the visit ratio values of the nodes are equivalent to PageRank values of the nodes.
For the {\em C\_VR} sample the states are sampled with probability inversely proportional to the visit ratio values
of the nodes corresponding to them. 

\section{Implementation details}\label{tab:details}
In this section we describe various complications that arose during the implementation
and how we addressed them.

{\em Fetching:}
In our walks we did not crawl the web pages whose encoded version were
more than $300$ characters long following~\cite{bar-yossef-et-al-toplas-00}.  
We only downloaded HTML/Text documents, ignored Javascript links and frame src links on them.
To avoid wasting bandwidth, we downloaded only the first $5$ MB of a web page.
We stopped fetching a web page if we could not download it after
$1,500$ seconds. In this case Walk~A and Walk B selected uniformly at random
a sibling of the current node, while Walk~C made a random reset.

{\em Host overload:}
If a walk tried to fetch web pages on the same host consecutively more than
$3,000$ times, we put the walk to sleep for $20$ minutes to avoid
host overload. If this happened $12$ times on the same host, we stopped the walk
and declared that it was unable to leave the host.

{\em  Parallel links:}
If there were multiple parallel hyperlinks from one web page to another, we kept only two of them.

{\em HTTP and HTML redirects:}
If a web page redirected to an another web page we treated them as the same node in the web graph. 
This applied iteratively to the whole ``redirect chain''. We combined the inlinks of
all the web pages in the redirect chain. If this combination resulted in more than 10 inlinks retrieved
from a search engine, we stored only a uniform random sample of 10 of them. If a newly visited web page redirected
to a previously visited web page, we did not retrieve inlinks for the new web page and instead used the inlinks
of the previous web page.
We followed only up to 10 HTTP or HTML redirects. If there were more than 10 redirects
or we detected a redirection loop, Walk~A and Walk B selected a random sibling of the previous 
node while Walk~C made a random reset.

{\em Truncation:} URLs with and without session id usually represent the same web page.
Thus, we treated them as one node to avoid bias during the walk
and the subsampling phase. Session ids are notoriously hard to detect in general, but frequently
they come after question marks in the URLs of the web pages. Thus we truncated URLs with
question mark at the question mark, but only under certain conditions.
First we experimented with a walk that always truncated at session ids.
However, sites for webmaster referral programs frequently encode a web page
after the question mark and redirect to it.
Truncating after the question mark prevented the walk to follow those redirections.
Truncating only if no error page is returned does not solve the problem either because
the truncated page might not return an error page but cause a new redirection.
Thus we chose the following strategy: When fetching a web page
the walk first follows all redirects that it can follow and
if the URL of the final web page in the redirect chain contains a question mark it is truncated.
If the truncation leads to an error page or a new HTTP redirect, the walk undoes the truncation.

{\em Speed up:}\label{tab:speedup}
To speed up the walk phase of all the algorithms we used multiple walks in parallel which shared the database. 
These walks started from the same initial node and they were not completely independent of each other since
they shared the database. However, the shared database only makes sure that all the walks
``see the same graph'', i.e., that the edges adjacent to a node remain same
throughout all the walks.

{\em Sampling steps or states:} 
In the subsampling phase the last half of the steps of the multiple walks 
are merged and a subgraph is formed from these steps.
We recorded the number of times the merged walk spent at each node, namely the {\em visit count} for each node.
For the A\_StatesOnLastHalf sample the states of the merged Walk~A on the formed subgraph are sampled with probability
inversely proportional to the degree of the states.
For the A\_StepsOnLastHalf sample the states of the formed subgraph are sampled with probability proportional to the state's visit count divided by its degree.
The samples from the last quarter of the steps of the multiple walks are taken exactly the same way 
except that we formed the subgraph from the last quarter of the steps of each walk. 
We proceded in the same way for the other algorithms. We set the sampling probabilities such that
each sample consisted of around $10,000$ nodes.
 
{\em Average of samples:} For each sample type of each algorithm we took 5 samples.
Each number given in Section~\ref{tab:experiments} is actually the average of these 5 samples.
 
\section{Experiments}\label{tab:experiments}

Recall that Walk~A and Walk~B differ only in the number of selfloops
in the underlying graphs on which they are performed. To save resources we did not perform
a random walk for Algorithm~A and a random walk for Algorithm~B.
Instead we performed only one random walk ignoring selfloop steps 
for Algorithm~A and Algorithm~B.
We call this {\em Walk~AB}. In a postprocessing step we simulated Walk~A and Walk B
with selfloop steps by flipping a suitably-biased random coin (dependent on the algorithm)
once at every step of Walk~AB and adding a suitable number of selfloop steps
when the coin comes up heads.
For Walk~B the probability of traversing a selfloop is very high.
Thus instead of flipping often a random coin each deciding on just one step, namely the next one, we
model the number of selfloop steps at the current node by a geometric
random variable and determine how many selfloop steps
are executed at the current node using one random number.
This approach was already proposed by~\cite{bar-yossef-et-al-toplas-00}.
It results in exactly the same random walk as Walk~A, resp. Walk B, would have performed
with the same coin flips and random walk choices.
As a result of simulating Walk~A and Walk~B from one common walk, the data for Walk~A and Walk~B are highly correlated. 
However this has the positive side-effect that it allows to evaluate whether Algorithm~A,
which is a modification of Algorithm~B, does lead to better results, as claimed
by~\cite{nec}. In our implementation Algorithm~A and Algorithm~B agree in all
non-selfloop transitions. Thus, if changing the number of selfloops per state and subsampling
states inversely proportionally to degree instead of randomly
does indeed change the quality of the sample as claimed by~\cite{nec} our evaluation
should show that. We performed a completely separate random walk for Algorithm~C.

We ran both walks, Walk~AB and Walk~C, for 240 hours on two identical machines equipped with a
Intel Pentium 4 processor 3.0 Hz (HyperThreading enabled), 4 GB of RAM, and a
4 Seagate HD (250GB each) in RAID5 on 3ware RAID controller 8506. As database
we used  PostgreSQL 7.4.8. The implementations shared as much code as possible.
Both Walk~AB and Walks~C started from {\em http://www.yahoo.com/} and used 50 walks in parallel
as explained in Section~\ref{tab:speedup}.
Three of the walks of Walk~AB had to be stopped because of host overloading 
before the end of the walk.
We removed their nodes and transitions from Walk~AB.
None of the C walks had to be stopped.
Walk~AB visited 842,685 nodes, leading to 1.7 million steps for Algorithm~A and 
4.3 trillion steps for Algorithm~B. 
Walk~C visited 695,458 nodes with almost 1 million steps. 

\begin{table}[htbp]
\vspace{-2.0mm}
\centering
\begin{tabular}{|l|l|l|l|r|r|}
\hline
Random walk&Duration& \# of visited nodes& \# of seen hosts & \# of seen domains\\
\hline
AB&$240 $ hours & $842,685$ &$2,360,329 $ & $1,041,903 $ \\
C&$240$ hours & $695,458 $  &$1,814,444 $ & $991,687 $ \\
\hline
\end{tabular}
\caption{Random walks on the web}
\label{tab:walks}
\vspace{-2.0mm}
\end{table}

Table~\ref{tab:walks} shows that the number of seen domains is almost identical
for Walk~AB and Walk~C. When compared to Walk~AB,
Walk~C visited 20\% fewer nodes and saw about 25\% fewer hosts.
This drop is not surprising since Walk~C made
a random jump to an already seen node in about 21\% of the transitions  
while Walk~AB does not perform random jumps.

In Walk~AB about 58\% of the non-selfloop transitions traversed an outlink, 42\% traversed an inlink.
We conjecture that the reason for this imbalance is that we artificially limit the number of inlinks
at 10, while the average number of outlinks for Walk~AB is $46.71$.

In Walk~C an outlink was traversed in
79\% and a random jump happened in 21\% of all the transitions. This number does not
vary much over the length of the walk. Based on the reset probability of $1/7$ one would
expect  that random jumps account only in 14\% of the transitions in Walk~C. However
dead ends, problems while fetching a page, long redirect chains, or redirect loops all caused
a random jump and are the reason for the additional 7\% of transitions with random jumps. 

Each of the following subsections compares Walk~A, Walk~B, Walk~C
and the samples generated by them using different measures.
The first subsection compares the algorithms using their ``nodes per host'' distribution.
The following subsections compare the algorithms using their ``PageRank bias'' and ``outdegree'' distribution.
These subsections all point out the weaknesses of different sampling approaches.
The last two subsection presents results for connectivity-independent statistics namely the
``top level domain'' and  the ``document content length'' distribution.

\subsection{Nodes per host distribution}
A uniform random sample of the nodes on the web graph should contain about
as many different hosts as there are nodes in the sample~\cite{bharat-et-al-toplas01}.
This is the case for each of C~Samples, each contain about 9,500
unique hosts out of about 10,000 nodes. However, A~Samples and B~Samples contain many fewer hosts
even though we omitted all the data from the three walks of Walk~AB that were stopped
because they were unable to leave a host.

\begin{table}[htbp]
\vspace{-1.8mm}
\centering
\begin{tabular}{|r|r|c|}
\hline
\# of nodes& \% of nodes & Host\\
\hline
\multicolumn{3}{|c|}{ A\_StatesOnLastHalf}\\
\hline
2051 &20.60\% &fr.shopping.com \\
874 & 8.78\% & www.rechtschutzversicherung.de \\
520 & 5.22\% & www.friday.littledusty.org \\
\hline
\multicolumn{3}{|c|}{ A\_StatesOnLastQuarter}\\
\hline
1850 & 18.94\%& fr.shopping.com \\
648 & 6.63\% & www.rechtschutzversicherung.de\\
627 &6.41\% & classifieds.fr \\
\hline
\multicolumn{3}{|c|}{ A\_StepsOnLastHalf}\\
\hline
2849 & 29.49\% &fr.shopping.com \\
874 &9.04\% &www.hostpooling.com \\
771 & 7.98\% &www.friday.littledusty.org \\
\hline
\multicolumn{3}{|c|}{ A\_StepsOnLastQuarter}\\
\hline
3170 & 32.73\% & fr.shopping.com\\
825 &  8.44\% &www.hostpooling.com  \\
677 & 6.92& www.friday.littledusty.org\\
\hline
\multicolumn{3}{|c|}{ B\_StatesOnLastHalf}\\
\hline
916 &9.11\% &fr.shopping.com \\
455 &4.53\% & www.rechtschutzversicherung.de \\
356 &3.54\% &  www.smart.com\\
\hline
\multicolumn{3}{|c|}{ B\_StatesOnLastQuarter}\\
\hline
730 &7.34\% & fr.shopping.com\\
302 &3.03\% & www.rechtschutzversicherung.de\\
256 &2.57\% & www.smart.com \\
\hline
\multicolumn{3}{|c|}{B\_StepsOnLastHalf}\\
\hline
2551 &26.63\% &fr.shopping.com\\
880 &9.18\% & www.hostpooling.com \\
521 &5.44\% & www.friday.littledusty.org\\
\hline
\multicolumn{3}{|c|}{B\_StepsOnLastQuarter}\\
\hline
2787 &29.13\% & fr.shopping.com \\
833 & 8.70\% & www.hostpooling.com \\
542 &5.67\% &classifieds.fr \\
\hline
\end{tabular}
\caption{The hosts with the most nodes in A~Samples and B~Samples}
\label{tab:BsamplesTopHosts}
\vspace{-1.8mm}
\end{table}

\begin{table}[htbp]
\centering
\begin{tabular}{|l|r|r|}
\hline
Sample &  \# of unique hosts\\
\hline
A\_StatesOnLastHalf &1,449\\
A\_StatesOnLastQuarter &1,277\\
A\_StepsOnLastHalf & 671\\
A\_StepsOnLastQuarter & 702\\
\hline
B\_StatesOnLastHalf &3,405  \\
B\_StatesOnLastQuarter &3,442\\
B\_StepsOnLastHalf & 656\\
B\_StepsOnLastQuarter & 750\\
\hline
C\_VR  &9,498\\
C\_PR  & 9,504\\
C\_Random &9,499 \\
\hline
\end{tabular}
\caption{The number of unique hosts in A~Samples, B~Samples and C~Samples}
\label{tab:Host count}
\end{table}

As can be seen in Table~\ref{tab:BsamplesTopHosts}, all A~Samples and B~Samples
except the B~samples subsampling states
contain about three times as many visited nodes on the host {\em fr.shopping.com}
than from other hosts. This is a significant bias towards the nodes on that host being 
due to multiple walks almost ``getting stuck'' in it.
It seems like a fundamental flaw in Algorithm~A and Algorithm~B: They have a large problem with hosts
that are highly connected within but have few edges leaving them. 
Here is an intuitive explanation: 
Consider an undirected graph of $n$ nodes, consisting of a complete graph of $n/2$ nodes 
with a chain of $n/2$ nodes attached to
one of the nodes in the complete graph. If Walk~A were run
on this graph, it would have a very good chance of getting stuck in the complete
subgraph when run long enough. To avoid this problem Algorithm~B 
added selfloops to make the graph regular. As a
result the walk is equally likely to ``get stuck'' on the chain as in
the complete subgraph. However, the fundamental problem of ``getting
stuck'', i.e., staying within a small part of the graph,
is not solved. 
Walk~C avoids this problem by performing random jumps. Indeed, the host with the largest number
of states in any of C~Samples, {\em www.amazon.com}, has only 32 nodes in the sample.

Table~\ref{tab:BsamplesTopHosts} shows that subsampling from the last half of the steps or the 
last quarter of the steps does not seem to have a impact on the resulting samples for Algorithm~A and Algorithm~B. 
The top 2 hosts with the most nodes are same in A~Samples and B~Samples subsampling states. 
Subsampling from the last half of the steps or the last quarter of the steps does not affect
the top 3 hosts list for A~Samples and B~Samples subsampling steps either.
On the other hand subsampling states or steps seems to make a difference.
With no exception {\em fr.shopping.com} is the host
with the most nodes in A~Samples and B~Samples. However in the B\_StatesOnLastHalf sample
and in the B\_StatesOnLastQuarter sample the percentage of nodes on the top hosts is smaller when
compared to the other A~Samples and B~Samples.
Table~\ref{tab:Host count} presents the number of unique hosts in A~Samples, B~Samples and C~Samples.
It shows that the number of unique hosts in B~Samples subsampling states is almost 5 times greater than the number
for B~Samples subsampling steps. 
For C~Samples as can be observed in Table~\ref{tab:BsamplesTopHosts}
the number of unique hosts is roughly same as the number of the nodes.
Thus we can conclude that sampling states leads to less biased distribution for the number of nodes per host.

\subsection{PageRank bias}
{\em Page-based analysis:}
Walk~C tries to visit nodes roughly according to their PageRank values.
Thus the most frequently visited nodes should have high PageRank values.
Table~\ref{tab:compaqtopurls2005} presents the top visited 10 nodes\footnote{The most visited node is 
a web page on a tracking site for website visitors. This web page is
the result of our truncation of URLs after questions marks for many different web pages, i.e.,
it is an artifact of our implementation of session id handling.} during Walk~C.
\begin{table}[htbp]
\centering
\small
\begin{tabular}{|r|r|l|}
\hline
PRank&
\begin{minipage}{1cm}\centering\vspace*{1mm}Visit\\[-1ex]count\vspace*{1mm}\end{minipage}
&Node\\
\hline
8&929 &http://extreme-dm.com/tracking/\\
10&810 &http://www.google.com/\\
8&696 &http://www.macromedia.com/shockwave/download/download.cgi\\
-&478 &http://www.sitemeter.com/default.asp\\
10&364 &http://www.statcounter.com/\\
7&336 &http://www.mapquest.com/features/main.adp\\
10&312 &http://www.microsoft.com/windows/ie/default.mspx\\
9&312 &http://www.yahoo.com/\\
10&294 &http://www.adobe.com/products/acrobat/readstep2.html\\
9&286 &http://www.blogger.com/start\\
\hline
\end{tabular}\\
\caption{The most visited 10 nodes of our Walk~C}
\label{tab:compaqtopurls2005}
\end{table}
We also give the PageRank as returned by the Google toolbar next to each node.
For one node no PageRank
is returned, all others have Toolbar PageRank 7 or above. We conclude that
our walk did indeed succeed in
visiting high PageRank nodes more frequently than other nodes.
We observed no such bias towards high PageRank nodes in Walk~AB
as can be seen from Table~\ref{tab:WalkABtopurls}.
Indeed no PageRank value is returned by the Google toolbar.

\begin{table}[htbp]
\centering
\small
\begin{tabular}{|r|r|l|}
\hline
PRank&
\begin{minipage}{1cm}\centering\vspace*{1mm}Visit\\[-1ex]count\vspace*{1mm}\end{minipage}
&Node\\
\hline
-&10,228&http://66.40.10.184/browses/AlphaBrowses/NF\_manufacturer.asp\\
-&7,496&http://www.mix-networks.com/forum/index.php\\
-&7,436&http://www.fatmp3.com/sitemap.html\\
-&6,899&http://bbs.dingding.org/RssFeed.asp\\
-&5,005&http://www.hotels55.info/a-z-test.php\\
-&2,457&http://www.hostpooling.com/berlin/hotel/billig/lease/home\_\&\_garden.htm\\
-&2,434&http://sms.3721.com/rsearch/ivr.htm\\
-&2,411&http://www.sh-netsail.com/www7/default.asp\\
-&2,185&http://www.hostpooling.com/berlin/hotel/billig/lease/health.htm\\
-&1,999&http://forums.gamedaily.com/index.php\\
\hline
\end{tabular}\\
\caption[]{The most visited 10 nodes of Walk~AB}
\label{tab:WalkABtopurls}
\end{table}

\begin{figure}[htbp]
\begin{minipage}[b]{0.50\textwidth}
\centering\includegraphics[width = \textwidth]{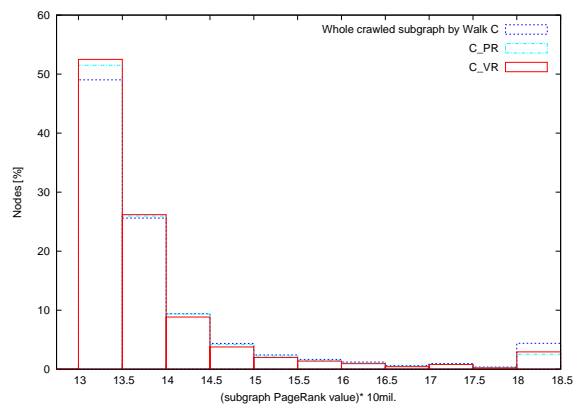}
\end{minipage}
\caption{PageRank value distribution in the crawled subgraph for Walk~C, in C\_PR sample and in C\_VR sample}
\label{fig:pagerank}
\vspace*{-3mm}
\end{figure}

We call the PageRank of the subgraph traversed in Walk~C the {\em subgraph PageRank}\footnote{The subgraph 
PageRank value of a state can be very different from its PageRank value in the whole web graph.}.
Figure~\ref{fig:pagerank} shows the percentage of nodes in certain subgraph PageRank ranges
for the whole crawled subgraph, for the C\_PR sample and for the C\_VR sample.
Since the C\_PR sample was created by subsampling states inversely proportional to
the subgraph PageRank values we would expect that nodes
with low subgraph PageRank values are more frequent in the sample than in the graph
as a whole and very few nodes with high subgraph PageRank values are in the sample. This
is exactly what we see in Figure~\ref{fig:pagerank}.
We also included the C\_VR sample in the figure although we did not use PageRank values
for getting the C\_VR sample. In Figure~\ref{fig:pagerank} we see that the C\_VR sample behaves
very similar to the C\_PR Sample. This shows that using the visit ratio as a substitute to PageRank
works as well. However neither subsampling phase is powerful enough to erase the PageRank bias present in Walk~C.

{\em Host-based analysis:}
The visit count of a node is the number of visits to the node as defined in Section~\ref{tab:Description}.
The visit count of a host is the sum of the visit counts of the nodes, namely web pages, on that host.
Table~\ref{tab:compaqtophosts} shows the top visited hosts of Walk~C
together with their visit counts. It shows a clear bias towards well-known popular
hosts. The most visited 10 hosts of Walk~AB show no obvious bias to well-known, popular hosts.
Table~\ref{tab:compaqtophosts}
shows also the top visited hosts of Walk 2 in~\cite{henzinger-et-al-toplas99}.
Only three of the hosts, namely Amazon, Microsoft and Adobe,
are in the top 10 list for both years. We attribute these differences to the big changes that have occurred
in the web in the mean time.

\begin{table}[htbp]
\centering
\begin{tabular}{|r|l|l|r|}
\hline
\multicolumn{2}{|c|}{Our Walk~C} & \multicolumn{2}{|c|}{ Walk~2 in~\cite{henzinger-et-al-toplas99}}\\
\hline
Visit count  &Host &Host  &Visit count \\
\hline
4,509&www.macromedia.com  &www.microsoft.com&32,452\\
3,262&www.amazon.com&home.netscape.com&23,329\\
2,848&www.google.com  &www.adobe.com&10,884\\
2,246&www.microsoft.com  &www.amazon.com&10,146\\
1,617&www.cyberpatrol.com&www.netscape.com&4,862\\
1,462&www.sedo.com   &excite.netscape.com&4,714\\
1,412&www.adobe.com    &www.real.com&4,494\\
1,132&www.cafepress.com&www.lycos.com&4,448\\
1,069&www.blogger.com    &www.zdnet.com&4,038\\
929&extreme-dm.com  &www.linkexchange.com&3,738\\
\hline
\end{tabular}\\
\caption[]{The most visited 10 hosts of our Walk~C and of Walk 2 in~\cite{henzinger-et-al-toplas99}}
\label{tab:compaqtophosts}
\end{table}
\subsection{Outdegree distribution}

\begin{figure}[htbp]
\subfigure[]{
\begin{minipage}[b]{.45\textwidth}
\centering\includegraphics[width = \textwidth]{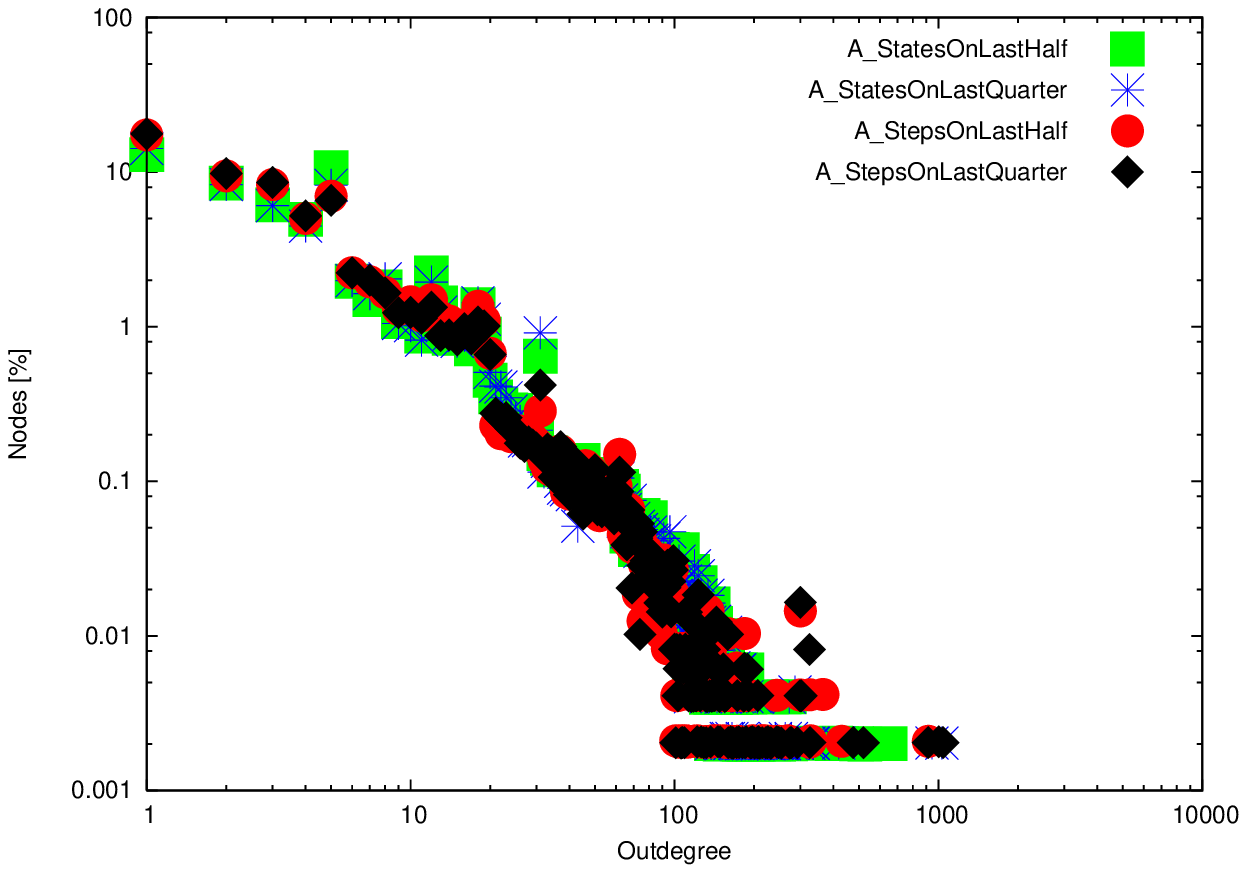}
\label{fig:outdegree_nec_log}	
\end{minipage}}
\subfigure[]{
\begin{minipage}[b]{.45\textwidth}
\centering\includegraphics[width = \textwidth]{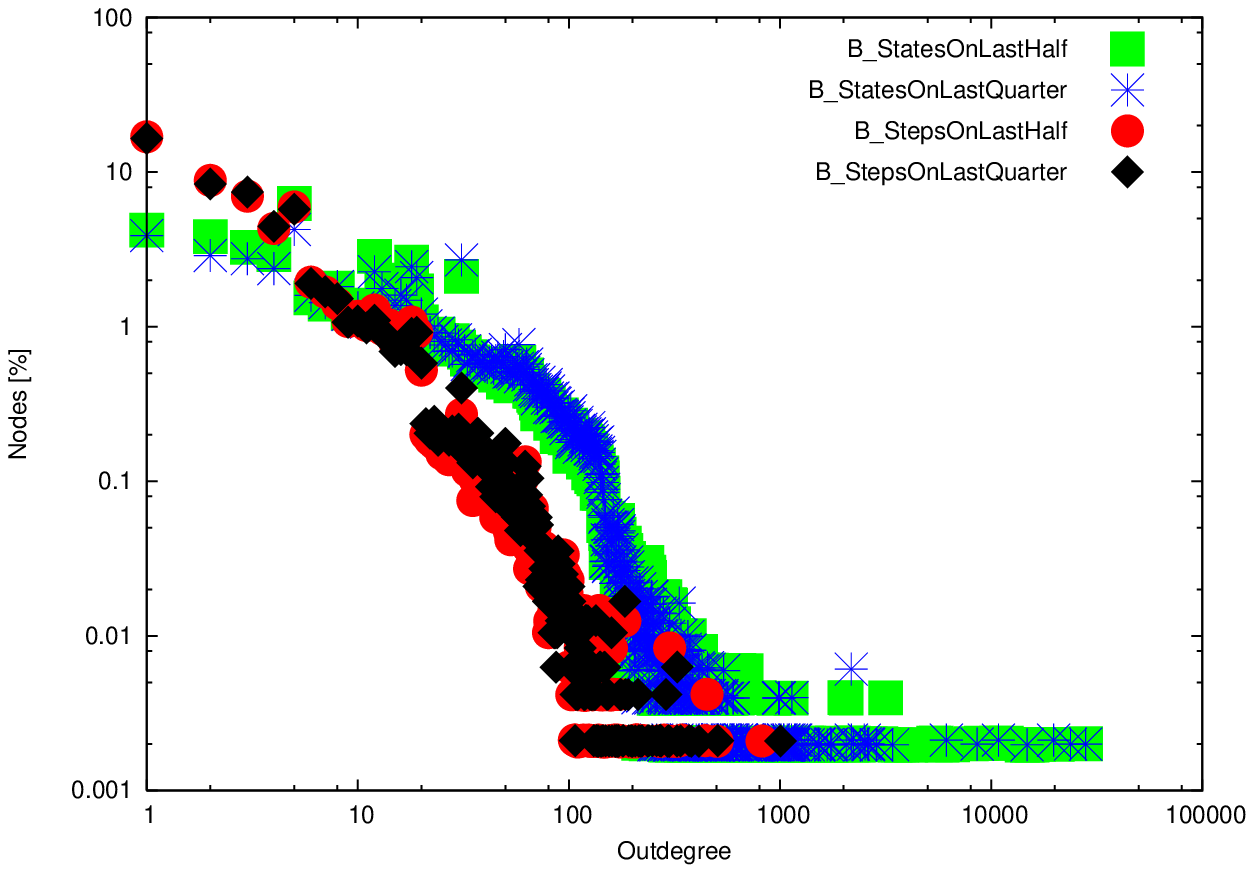}
\label{fig:outdegree_berkeley_log}
\end{minipage}}
\subfigure[]{
\begin{minipage}[b]{.45\textwidth}
\centering\includegraphics[width = \textwidth]{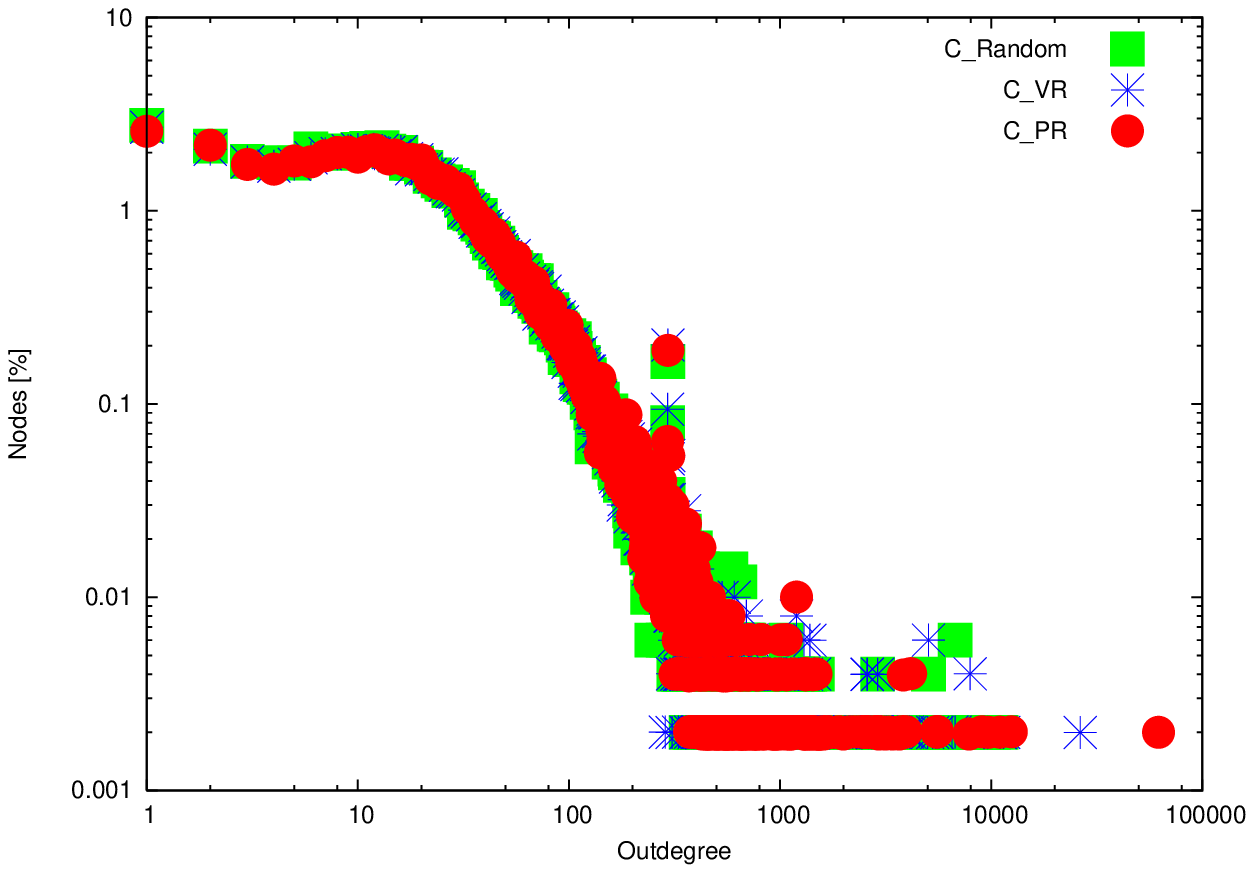}
\label{fig:outdegree_compaq_log}
\end{minipage}}
\caption{Outdegree distribution of nodes in (a) A~Samples (on log-log scale),
(b) B~Samples (on log-log scale), (c) C~Samples (on log-log scale)}
\label{fig:Outdegree}
\end{figure}

As was shown in the literature the outdegree distribution of the nodes on the web graph follows a power law. 
Thus the outdegree distribution of a uniform random sample of the nodes on the web graph
should ideally follow a power law distribution.
In Figure~\ref{fig:outdegree_nec_log}, Figure~\ref{fig:outdegree_berkeley_log} and Figure ~\ref{fig:outdegree_compaq_log} 
we present the outdegree distribution on log-log scale for A~Samples, B~Samples and C~Samples respectively. 
In these figures for all the samples we observe that the percentage of nodes
with high outdegree is lower when compared to the percentage of nodes with low outdegree. 

The power law exponent for outdegree distribution of the nodes on web graph is given as 2.72 in~\cite{broder-et-al-toplas00}.
It would be interesting to see how our samples agree with this value.
For the A\_StatesOnLastHalf sample and the B\_StepsOnLastHalf sample the power law exponent
is 2.01. On the other hand for the B\_StatesOnLastHalf sample the exponent is 1.41.
In other words, B~samples sampling states are more biased to high outdegree nodes when compared to other B~Samples and A~Samples.
The outdegree power law exponent is about 1.49 for C~Samples. This indicates that C~Samples have
a bias to high outdegree nodes.
For none of our samples does the power law exponent agree with the value in
the literature, giving evidence that all our samples are biased to high outdegree nodes.

\begin{table}[t]
\vspace{-2.0mm}
\centering
\begin{tabular}{|l|r|r|}
\hline
Sample&\multicolumn{2}{c|}{Outdegree}\\
\hline
 & Avg & Max \\
\hline
A\_StatesOnLastHalf &8.03 & 656 \\
A\_StatesOnLastQuarter &7.80 &1,031 \\
A\_StepsOnLastHalf &7.01 &916\\
A\_StepsOnLastQuarter &6.95 &1,041\\
\hline
B\_StatesOnLastHalf &46.76 & 27,994\\
B\_StatesOnLastQuarter &47.08 &27,994\\
B\_StepsOnLastHalf &6.63 & 822\\
B\_StepsOnLastQuarter &6.63 &1,003\\
\hline
C\_VR &59.06 &26,423\\
C\_PR &60.22 &62,021\\
C\_Random &57.82 &11,138 \\
\hline
\end{tabular}
\caption{Statistics about outdegree distribution}
\label{tab:StatisticsOutDegree}
\vspace{-2.0mm}
\end{table}

The average outdegree on the web graph was estimated by prior work~\cite{kleinberg-et-al-toplas99}
to be around $10$. A uniform random sample of the web graph should have this property. In order
to investigate this we present the statistics about outdegree distribution of A~Samples, B~Samples and C~Samples
in Table~\ref{tab:StatisticsOutDegree}.
As can be seen in this table all A~Samples, the B\_StepsOnLastHalf sample and the B\_StepsOnLastQuarter sample 
have an average outdegree of roughly 10. However, the C~Samples,
the B\_StatesOnLastHalf sample and the B\_StatesOnLastQuarter sample have an average outdegree 
that is a factor of 4 to 6 larger. We already discussed above that the C~Samples have a bias to high outdegree nodes.
At a first glance looking at the average outdegree seems to indicate that the A
and the B samples sampling steps have no bias towards high outdegree nodes, contradicting our above finding based on
the power law exponent.
However a closer investigation 
showed that many of the most frequently visited hosts have nodes with outdegree 0.
Thus the very biased distribution of nodes per host of the samples
sampling steps leads to their low average outdegree.

\newpage
\subsection{Top level domain (TLD) distribution}
A top level domain is the last part of the domain name, like
``.com'' or ``.net''. The distribution of web pages
over the top level domains is not known, but could be estimated if we
could sample the web uniformly at random. Even though, unlike for the outdegree distribution or the
nodes per host distribution we do not know the ``correct'' answer, it is interesting to
compare the results achieved by the different sampling techniques. A rough
agreement would give us an indication of what the correct answer is likely to be.
Thus in this subsection we present the 
top level domain distribution for A~Samples, B~Samples and C~Samples (see Table~\ref{tab:tld_undirected}
and Table~\ref{tab:tld_directed}).
\begin{table}[htbp]
\centering
\begin{tabular}{|l|r|r|r|r|r|r|r|r|}
\hline
 TLD& \multicolumn{4}{c|}{A~Samples} & \multicolumn{4}{c|}{B~Samples} \\
\hline
& {A\_States}& {A\_States}  & {A\_Steps} & {A\_Steps} & {B\_States}& {B\_States}  & {B\_Steps} &{B\_Steps}  \\
& {OnLast}& {OnLast} & {OnLast} & {OnLast}& {OnLast}& {OnLast} & {OnLast} &{OnLast} \\
& {Half}& {Quarter} & {Half} & {Quarter}& {Half}& {Quarter} & {Half} &{Quarter} \\
\hline
.com& 53.81& 50.82 & 64.87 &62.77& 49.29& 44.26 & 64.55 & 62.67 \\
.edu& 0.22& 0.24 & 0.06 &0.08& 0.41& 0.32 & 0.05 & 0.08 \\
.org& 8.26& 8.67 & 12.57 &11.72& 4.18& 4.23 & 10.89 & 8.70 \\
.net& 6.83& 7.39 & 8.42 &7.71& 8.93& 10.11 & 8.17 & 7.85 \\
.jp& 0.81& 0.96 & 0.25 &0.38& 2.09& 2.62 & 0.24 & 0.39 \\
.gov& 0.13& 0.15& 0.05 &0.03& 0.24& 0.20 & 0.05 & 0.04 \\
.uk& 1.26& 0.66 & 0.51 &0.29& 1.40& 0.99 & 0.59 & 0.33 \\
.us& 0.13& 0.16 & 1.02 &1.29& 0.39& 0.46 & 0.97 & 1.87 \\
.de& 11.96& 11.02 & 3.70 &3.92& 7.74& 6.03 & 3.49 & 3.21 \\
.ca& 0.16& 0.12 & 0.04 &0.06& 0.33& 0.26 & 0.04 & 0.05 \\
.fr& 5.68& 8.44 & 2.32 &4.36& 1.53& 2.00 & 3.56 & 7.10 \\
\hline
\end{tabular}
\caption{Top level domain distribution for A~Samples and B~Samples}
\label{tab:tld_undirected}
\end{table}

\begin{table}[htbp]
\centering
\begin{tabular}{|l|r|r|r|r|r|}
\hline
TLD & \multicolumn{3}{c|}{C~Samples} & \multicolumn{2}{c|}{Samples from 2000} \\
\hline
 & {C\_Random} &{C\_PR} & {C\_VR} &{B Sample} & {C\_VR}\\
 & & &  & from~\cite{bar-yossef-et-al-toplas-00} & from~\cite{henzinger-et-al-toplas00}\\
\hline
.com&63.20&62.94&63.13& 49.15 &45.62 \\
.edu&0.64&0.60&0.67& 8.28 & 9.84\\
.org&9.79&9.94&9.82& 6.55 & 9.12\\
.net&6.19&6.14&6.20&5.60 & 4.74 \\
.jp&0.44&0.48&0.46& 2.87 & 3.87\\
.gov&0.47&0.46&0.49& 2.08 & 3.42 \\
.uk&3.28&3.34&3.26& 2.75 & 2.59\\
.us&0.63&0.62&0.56& 1.12 & 1.77\\
.de&3.28&3.32&3.28& 3.67 & 3.26\\
.ca&0.83&0.83&0.84& 1.58 & 2.05 \\
.fr&0.43&0.40&0.43& 1.01 & 0.99\\
\hline
\end{tabular}
\caption[]{Top level domain distribution for C~Samples, B Sample from~\cite{bar-yossef-et-al-toplas-00} 
and C\_VR sample from~\cite{henzinger-et-al-toplas00}}
\label{tab:tld_directed}
\end{table}

Recall that Walk~AB and Walk~C were performed completely independent of
each other. Still the samples generated from them roughly agree: About 44-65\%
of the nodes, namely web pages, are in ``.com'' domain, making it clearly the largest
domain on the web. 
The domains ``.net '' and ``.org'' contain 
about 4-9\% of the nodes. 

The domains ``.de'' and ``.fr'' show large
variances in the percentage of the nodes in them. For ``.de'' the large values (around 11\%)
for the A\_StatesOnLastHalf sample and the A\_StatesOnLastQuarter sample are due to the high frequency
of a German host, which in turn is caused by the inability of Walk~AB
of leaving highly connected hosts. Thus these percentages are artificially high and should be ignored.
Additionally all percentages for the ``.de'' domain are inflated due to the fact that we
performed our walks from Switzerland for which the country of originator
for domain forwarding is Germany. 

The results for top level domain distribution from~\cite{bar-yossef-et-al-toplas-00} and
from~\cite{henzinger-et-al-toplas00} (Table~\ref{tab:tld_directed}) roughly agree and ``.com'' is the largest
top level domain as in our A~Samples, B~Samples and C~Samples.

\subsection{Document content length distribution}
\begin{figure}[htbp]
\subfigure[]{
\begin{minipage}[b]{.45\textwidth}
\centering\includegraphics[width = \textwidth]{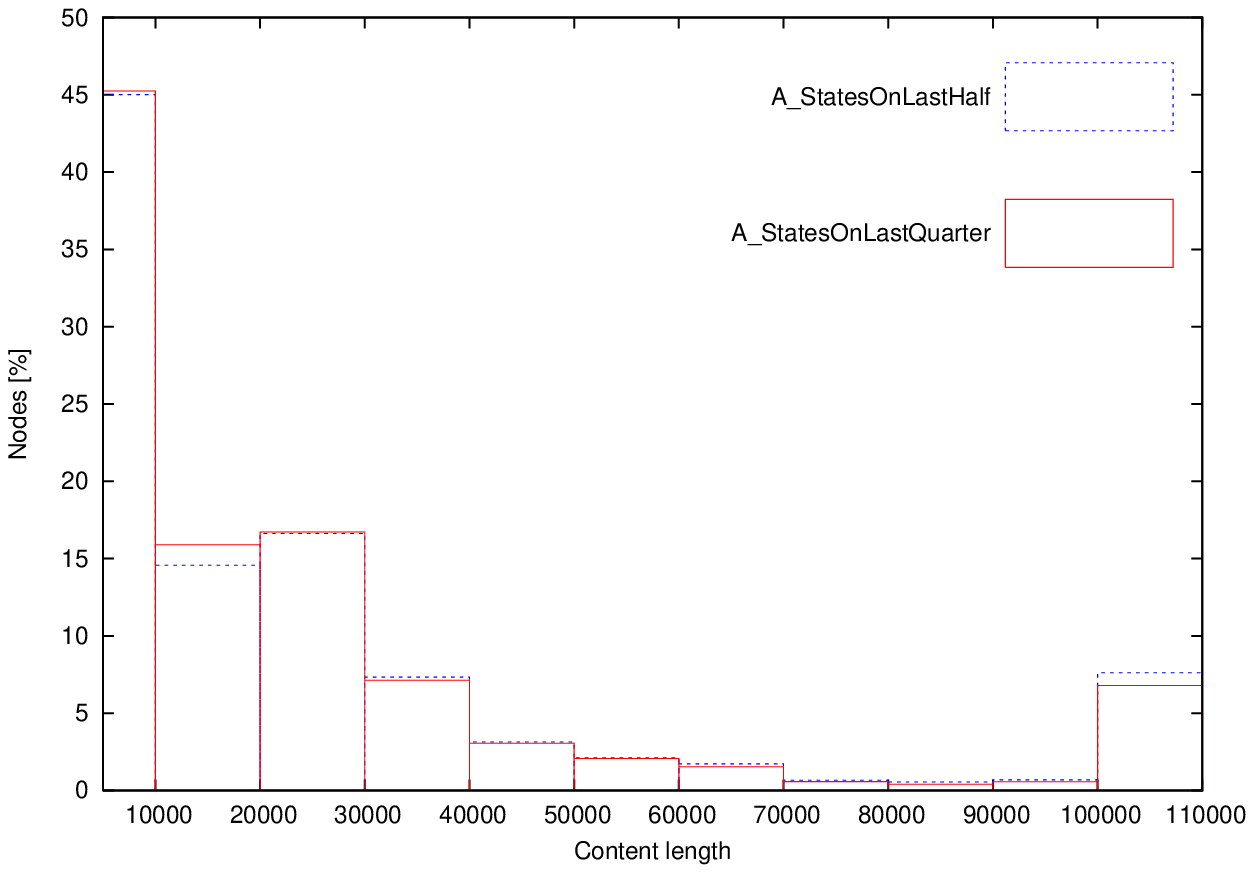}
\label{fig:content_nec_samplestates}
\end{minipage}}
\subfigure[]{
\begin{minipage}[b]{.45\textwidth}
\centering\includegraphics[width = \textwidth]{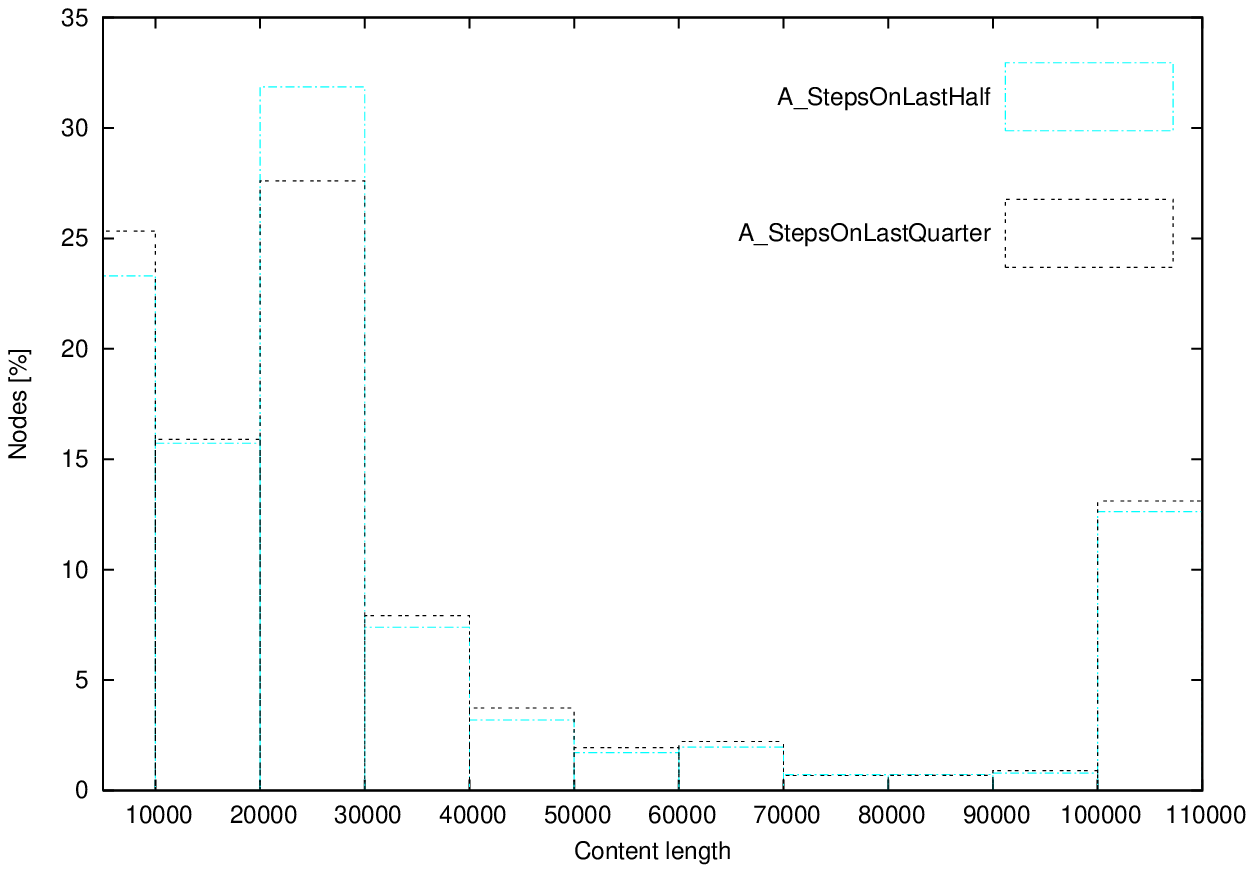}
 \label{fig:content_nec_samplesteps}
\end{minipage}}
\subfigure[]{
\begin{minipage}[b]{.45\textwidth}
\centering\includegraphics[width = \textwidth]{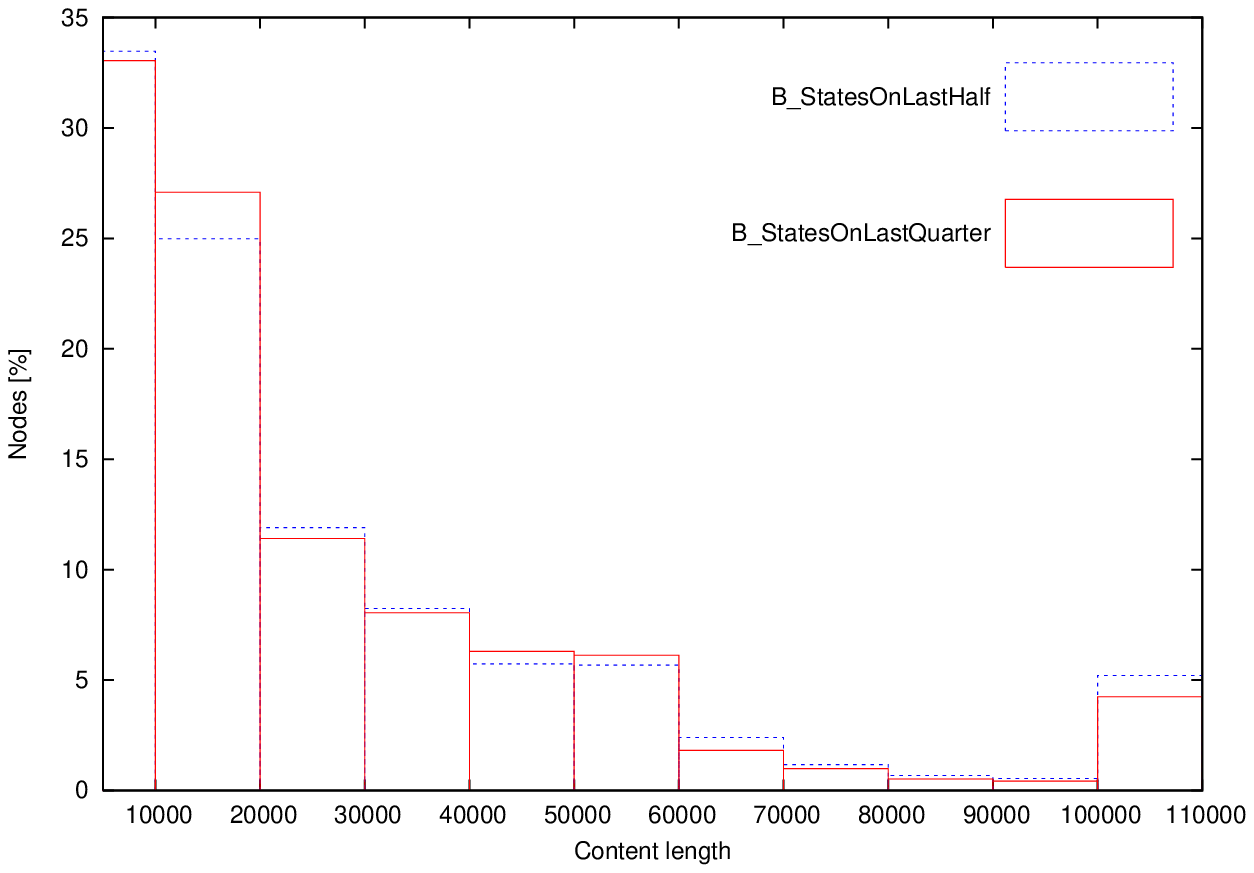}
 \label{fig:content_berkeley_samplestates}
\end{minipage}}
\subfigure[]{
\begin{minipage}[b]{.45\textwidth}
\centering\includegraphics[width = \textwidth]{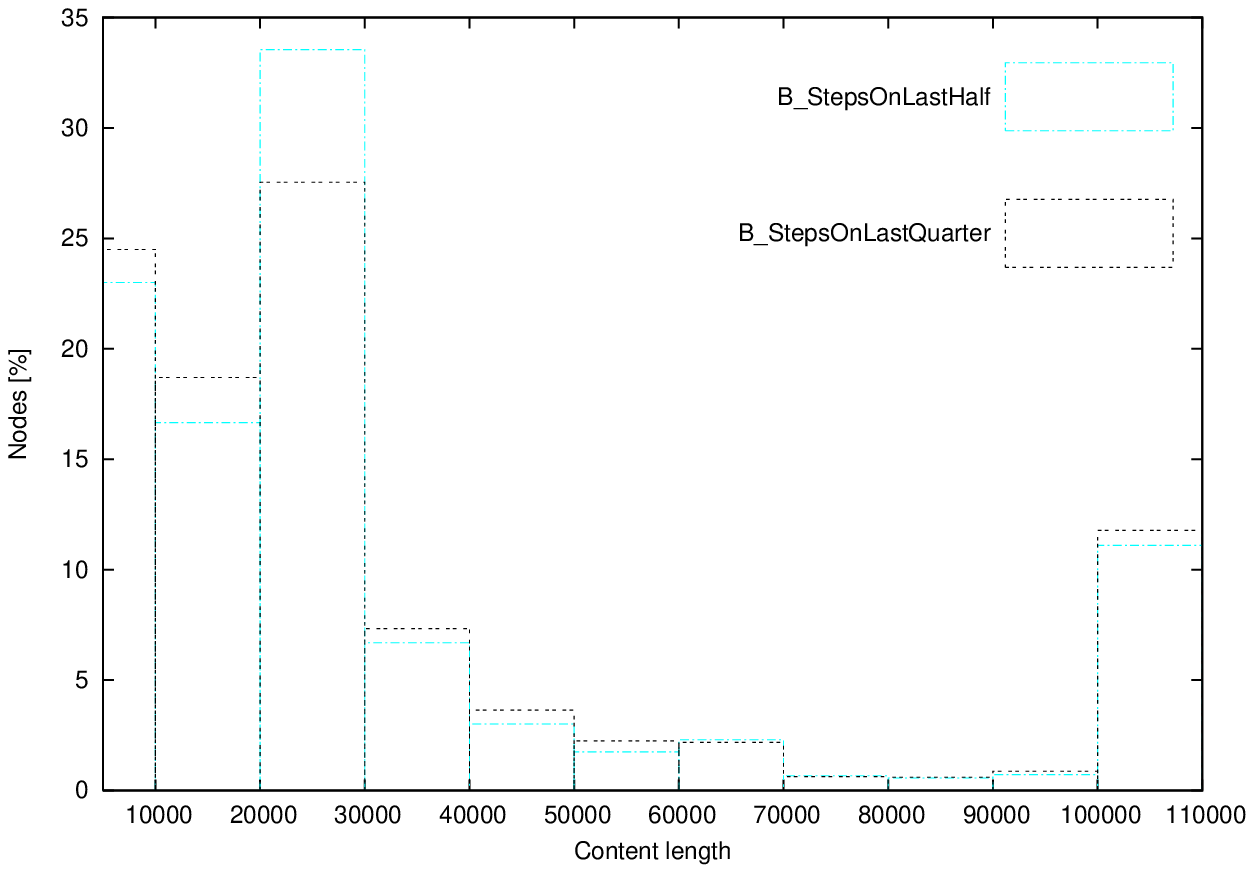}
\label{fig:content_berkeley_samplesteps}
\end{minipage}}
\subfigure[]{
\begin{minipage}[b]{.45\textwidth}
\centering\includegraphics[width = \textwidth]{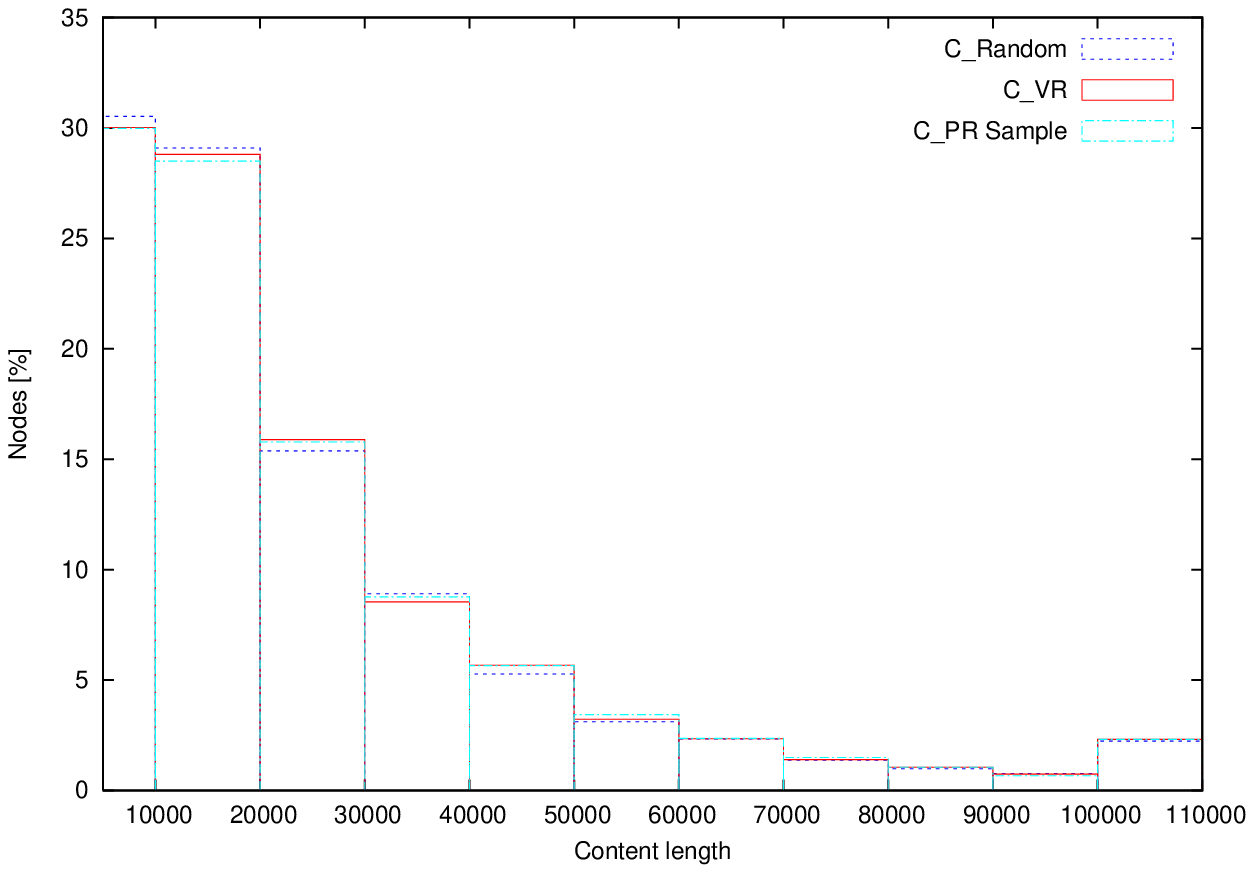}
\label{fig:ContentC}
\end{minipage}}
\caption{Content length (in bytes) distribution of nodes (web pages) in (a) A~Samples subsampling states, 
(b) A~Samples subsampling steps, 
(c) B~Samples subsampling states and (d) B~Samples subsampling steps (e) C~Samples}
\label{fig:ContentAB}
\end{figure}

In this subsection we study the document content length distribution for A~Samples, B~Samples and C~Samples.
We bucketed the content length values as follows: the first bucket, 0-10k, contains
the percentage of nodes (web pages) in the samples whose content length is between 0 and 10k.
The definition for the other buckets is analogous.
For the last bucket (100-110k) we put all the nodes whose content length is greater than 100k,
causing a relatively large value in that bucket for all the samples.

Figure~\ref{fig:ContentAB} presents the document content length distribution for the different samples.
B~Samples subsampling states (Figure~\ref{fig:content_berkeley_samplestates}) have a similar document content length distribution 
as C~Samples (Figure~\ref{fig:ContentC}).
Generally the percentage of nodes per bucket is monotonically decreasing with the content length.
However, there is a spike for A samples sampling states in bucket 0-10k and a spike for A
samples and B samples sampling steps in bucket 20-30k. A detailed analysis showed that
these spikes are caused by the uneven distribution of nodes over hosts.

\section{Comparison of techniques} \label{tab:summary}
In this section we compare the different samples of each algorithm over all the different measures we have used.

{\em Subsampling from the last half or from the last quarter of the steps}:
Since we ran walks for a fixed amount of time starting from the same node, the results
are somewhat influenced by the chosen starting node. The longer the walks run, the smaller
this bias should become. Thus we wanted to evaluate whether subsampling Walk~A and Walk B from the
last quarter of the steps gives improved results over subsampling these walks from the last half of the steps.
Our results indicate that this is not the case. For none of the samples
did we see a large difference in the results whether they were based on the last half or the last quarter of the steps. 
Thus either approach seems to work equally well and the starting-node bias seems small.

{\em Subsampling from steps versus from states}:
After determining the set of steps to subsample from, we either subsampled steps directly from these steps
or we determined the states represented by them and subsample the states. Obviously, when a random walk
was unable to leave a host for a long time and frequently revisits nodes on the same host, these nodes
have a higher chance of being in the sample when we subsample steps than when we subsample states. This can be
seen in Table~\ref{tab:BsamplesTopHosts}: When steps are subsampled, a much larger percentage
of the samples belongs to the same host  than when states are subsampled. 
As a result various measures exhibit unexpected spikes for the samples based on steps, see 
for example the document content length distribution in Figure~\ref{fig:ContentAB}. This indicates that it is better to subsample
states. 
However, for top-level domain distribution the samples based on steps both for Walk~A and
for Walk B showed a large agreement with C~Samples, while the samples based on states disagreed
with each other and with the C~Samples. Further investigation is necessary to understand this behaviour.

{\em Algorithm~A versus Algorithm~B versus Algorithm~C}:
Algorithm~C has a clear bias towards high PageRank and high outdegree nodes.
However, it generates a roughly uniform distribution of nodes per host. Algorithm~A
and Algorithm~B generate a very unbalanced distribution of nodes over hosts, with more
than 30\% of the nodes in the sample belonging to only three hosts. As a consequence
 it is hard to believe that the results
produced by this sample are representative of the whole web. All A~Samples as well as B~Samples
subsampling steps exhibit this problem. 
Thus Algorithm~B combined with state subsampling appears
superior to Algorithm~A. Recall that Algorithm~A and Algorithm~B were both implemented by
the same walk. They differ however, (1) by the number of selfloops of the nodes and (2) by the
subsampling probabilities (inversely proportionally to the degree for Algorithm~A and uniformly at random for Algorithm~B). Let us compare the A Samples subsampling states with the B Samples subsampling states.
Both subsample from the states in the last half or in the last quarter of the steps of the walk.
There are two possible reasons for the different quality of their samples: (1) Due to the selfloops the set of nodes from which Algorithm~A and Algorithm~B sample is very different. (2) Due to the
probabilities used for subsampling different nodes are picked.
To determine which of these reasons applies we compared the set of nodes 
used to subsample from. Our analysis showed they are almost identical for Algorithm~A and Algorithm~B.
Thus, the subsampling probabilities are the reason for the difference in host frequency distribution for A~Samples and B~Samples subsampling states.

\section{Conclusions and future work}\label{tab:futureWork}
We compared Algorithm~A, Algorithm~B and Algorithm~C under conditions
that are as equal as possible. Walk~C has a clear bias towards
high PageRank and high outdegree web pages and there seems to be no obvious way of correcting it.
Algorithm~A and Algorithm~B has a serious problem with ``getting stuck'' in hosts.
This had a clear impact on the nodes per host, outdegree, top level domain and document content length distribution.
However, we believe that this problem can be corrected.
We tried to eliminate the problem by stopping the walk when it could not leave a host for a large number
of steps. However, a better approach might be to perform a random reset every $x$
steps, like in Algorithm~C. This is also the approach taken by~\cite{yossef-et-al-focusedsampling} and by~\cite{chakrabarti-et-al-02} in their work on the distribution of topics on the web.
\bibliographystyle{plain}
\bibliography{references}

\end{document}